\documentclass{sf2a-conf2017}
\usepackage{graphicx}
\usepackage{hyperref}
\usepackage[]{natbib}  
\usepackage{epstopdf}
\usepackage{subcaption}
\usepackage{color}

\def\BibTeX{{\rm B\kern-.05em{\sc i\kern-.025em b}\kern-.08em
    T\kern-.1667em\lower.7ex\hbox{E}\kern-.125emX}}
\bibpunct{(}{)}{;}{a}{}{,}  

\begin{document}

\TitreGlobal{SF2A 2017}


\title{Active galactic nuclei in the era of the Imaging X-ray Polarimetry Explorer}
\runningtitle{AGN in the era of IXPE}

\author{F.~Marin}\address{Université de Strasbourg, CNRS, Observatoire astronomique de Strasbourg, UMR 7550, F-67000 Strasbourg, France}
\author{M.~C.~Weisskopf}\address{NASA Marshall Space Flight Center, Huntsville, Alabama, United States}

\setcounter{page}{237}

\maketitle

\begin{abstract}
In about four years, the National Aeronautics and Space Administration (NASA) will launch a small 
explorer mission named the Imaging X-ray Polarimetry Explorer (IXPE). IXPE is a satellite dedicated 
to the observation of X-ray polarization from bright astronomical sources in the 2 -- 8~keV energy 
range. Using Gas Pixel Detectors (GPD), the mission will allow for the first time to acquire X-ray 
polarimetric imaging and spectroscopy of about a hundred of sources during its first two years of 
operation. Among them are the most powerful sources of light in the Universe: active galactic nuclei 
(AGN). In this proceedings, we summarize the scientific exploration we aim to achieve in the field 
of AGN using IXPE, describing the main discoveries that this new generation of X-ray polarimeters will 
be able to make. Among these discoveries, we expect to detect indisputable signatures of strong gravity, 
quantifying the amount and importance of scattering on distant cold material onto the iron K$\alpha$ 
line observed at 6.4~keV. IXPE will also be able to probe the morphology of parsec-scale AGN regions, 
the magnetic field strength and direction in quasar jets, and, among the most important results, deliver 
an independent measurement of the spin of black holes.
\end{abstract}



\section{Introduction}
The study of cosmic polarization led to numerous discoveries in almost all astronomical fields. The first 
measurement of starlight polarization goes back to \citet{Hiltner1949}, who found that the light of distant 
stars is polarized as high as 12\% due to interstellar clouds. More importantly, the position angle of the 
polarization was found to be close to the galactic plane, opening the way for a deeper comprehension of 
the nature of the interstellar medium \citep{Davis1951}. The polarization of the Sun itself was measured 
by many astrophysicists, with one of the most spectacular and earliest discoveries being the observation of 
the Zeeman effect in Sun spots by \citet{Hale1908}; the author evaluating for the first time the 
solar magnetic field strength \citep{Salet1910}. Broadband polarization was recorded for all possible 
astronomical sources, since it is present in radiation from coherent astronomical sources (in, e.g., 
astrophysical masers) to incoherent sources such as the large radio lobes in active galaxies. However,
not all energy windows are available nowadays for a systematic exploration of astronomical polarization.

The X-ray band is a step behind in comparison to all other wavebands in this regard. The first X-ray 
polarization measurements go back to \citet{Tindo1970} in the case of solar flares. Extra-solar X-ray 
polarization observations were achieved by \citet{Novick1972}, targeting the Crab nebula and several 
other bright X-ray sources. The unique precision measurement of the X-ray polarization of the Crab Nebula 
(without pulsar contamination) was achieved by \citet{Weisskopf1978} thanks to the Eighth Orbiting Solar 
Observator (OSO-8) graphite crystal polarimeters. Since then, only a couple of attempts were made in the 
hard X-ray band \citep{Suarez2006,Dean2008,Chauvin2017}.

In this proceedings we present the Imaging X-ray Polarimetry Explorer (IXPE), a NASA-led spatial mission 
that will fly in 2021 and carry, for the first time, a set of imaging X-ray polarimeters. Dedicated to 
the study of high energy sources, IXPE will re-open the window of X-ray polarimetric observations. 
Focusing on the field of active galactic nuclei (AGN), we here summarize what IXPE will be able to achieve 
in the first years of operation.

\section{The IXPE mission}
IXPE is part of the NASA's Explorer Mission project and is led by the Principal Investigator, Dr. Martin C. 
Weisskopf. Within a cost cap of \$~188~M, IXPE will be comprised of three identical, grazing-incidence, 
X-ray mirror modules assemblies that will collimate radiation to three accompanying polarization sensitive 
Gas Pixel Detectors (GPD). The NASA/MSFC is producing the X-ray mirror modules while the Italian Space Agency 
is providing the GPD, making IXPE an international mission. Ball Aerospace is taking care of the spacecraft 
and the services of mission integration that are also included in the cost of IXPE. The satellite will be 
launched from Kwajelein Atoll into a 540~km circular orbit at almost 0$^\circ$ inclination. The envisioned 
launcher is a Pegasus XL vehicle that can carry a mass of 450~kg. 

The nominal lifetime of IXPE is two years and about one hundred of targets will be observed, including AGN, 
microquasars, pulsars (plus wind nebulae), magnetars, X-ray binaries, supernova remnants, and the Galactic center. 
Thanks to its GPDs \citep{Costa2001,Bellazzini2006,Bellazzini2007}, IXPE will be able to measure the spatial, 
spectral, timing, and polarization state of X-rays in the 2 -- 8~keV band. NASA officially selected IXPE on 
January the 3$^{\rm rd}$, 2017, among fourteen proposals. As the first dedicated X-ray polarimetry observatory,
IXPE will significantly enlarge the observational phase space, probing fundamental questions concerning high densities,
high temperatures, non thermal particle acceleration, strong magnetic and electric fields, and strong gravity.
Additional details about the mission can be found on-line at \textcolor{blue}{https://wwwastro.msfc.nasa.gov/ixpe/} 
and in \citet{Weisskopf2016}.

\section{Scientific goals in the field of AGN}
As stated above, IXPE will observe a large variety of X-ray sources. Among them are the bright cores of active galaxies.
The presence of accreting supermassive black holes can be inferred from the near-infrared to the X-ray domains but only 
X-ray polarization measurements can probe the geometry, composition, temperature and physics of matter at the smallest 
gravitational radii. In the following, we give three examples of the importance of IXPE in solving several key questions 
about AGN.

\subsection{Spin determination}

\begin{figure}[ht!]
\centering
  \includegraphics[trim = 0mm 0mm 0mm 0mm, clip, width=14cm]{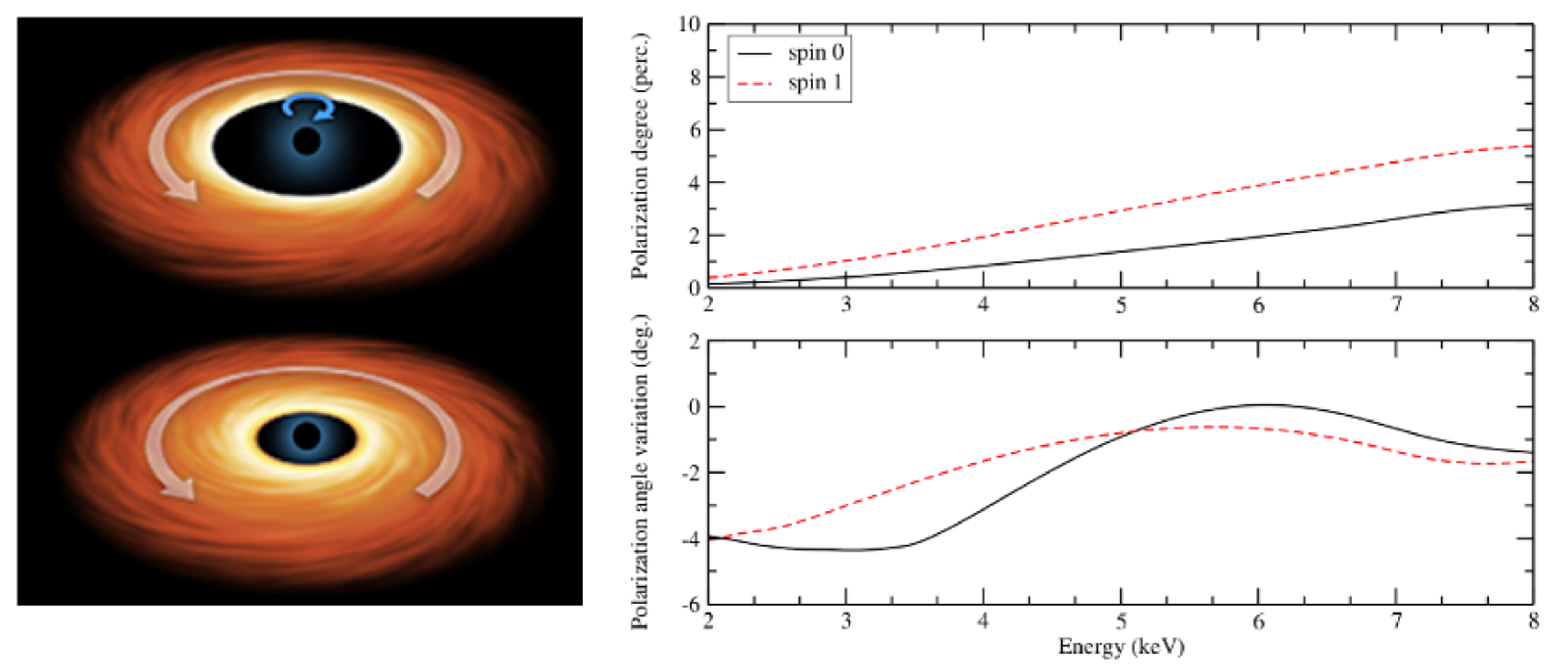}
  \caption{Spin determination through X-ray polarization continuum 
	  measurement. The left-hand panel represents two flavors 
	  of supermassive black hole spins: a non-rotating, a=0, 
	  Schwarzschild black hole on top and a maximally 
	  rotating, a=1, Kerr black hole below. Credits: 
	  NASA/JPL-Caltech. The right-hand panel is the 
	  polarization degree and polarization angle variation
	  associated with the two black hole spins in the IXPE
	  energy band. Simulations from \citet{Dovciak2011} 
	  and Marin et al. (2017, submitted).}
  \label{marin:fig1}
\end{figure}

Measuring the spin of black holes is a long-standing problem as the spin only affects the local spacetime around the 
potential well. It is thus necessary to observe the central parts of the accretion disk in the X-ray band in order
to determine the dimensionless angular momentum parameter, $a$, one of the two key parameters of black holes (with
their mass). Several methods exist: one can fit the profile of the relativistically-broadened iron K$\alpha$ line 
\citep[e.g.,][]{Reynolds2014} or fit the thermal X-ray continuum \citep[e.g.,][]{McClintock2014}. However the two 
methods do not always agree and there are many sources of possible systematic errors (e.g., intrinsic absorption, 
presence of a radio jet, modeling of the soft excess, role of emission from within the innermost stable circular orbit).

Measuring the X-ray polarization from AGN adds two independent parameters to the spectroscopic channel: the polarization 
degree and the polarization position angle. These new observational constraints remove many degrees of freedom from our 
current models that must fit both the spectroscopic and polarimetric data. In particular, it was shown by \citet{Schnittman2009} 
and \citet{Dovciak2011} that the X-ray polarization from X-ray binaries and AGN is particularly sensitive to the spin, luminosity
and inclination of the source. Fig.~\ref{marin:fig1} illustrates the difference between a non-spinning and a maximally spinning 
supermassive black hole in terms of polarization degree and angle\footnote{We show the rotation of the polarization position 
angle with respect to a convenient average of the polarization position angles over the depicted energy band. The actual 
normalization of the polarization angle with respect to the disk axis is not of primary interest as we cannot determine it 
from these observations \citep{Marin2012}.}.

\subsection{Strong gravity and distant scattering}

\begin{figure}[ht!]
 \centering
 \includegraphics[trim = 0mm 0mm 0mm 0mm, clip, width=14cm]{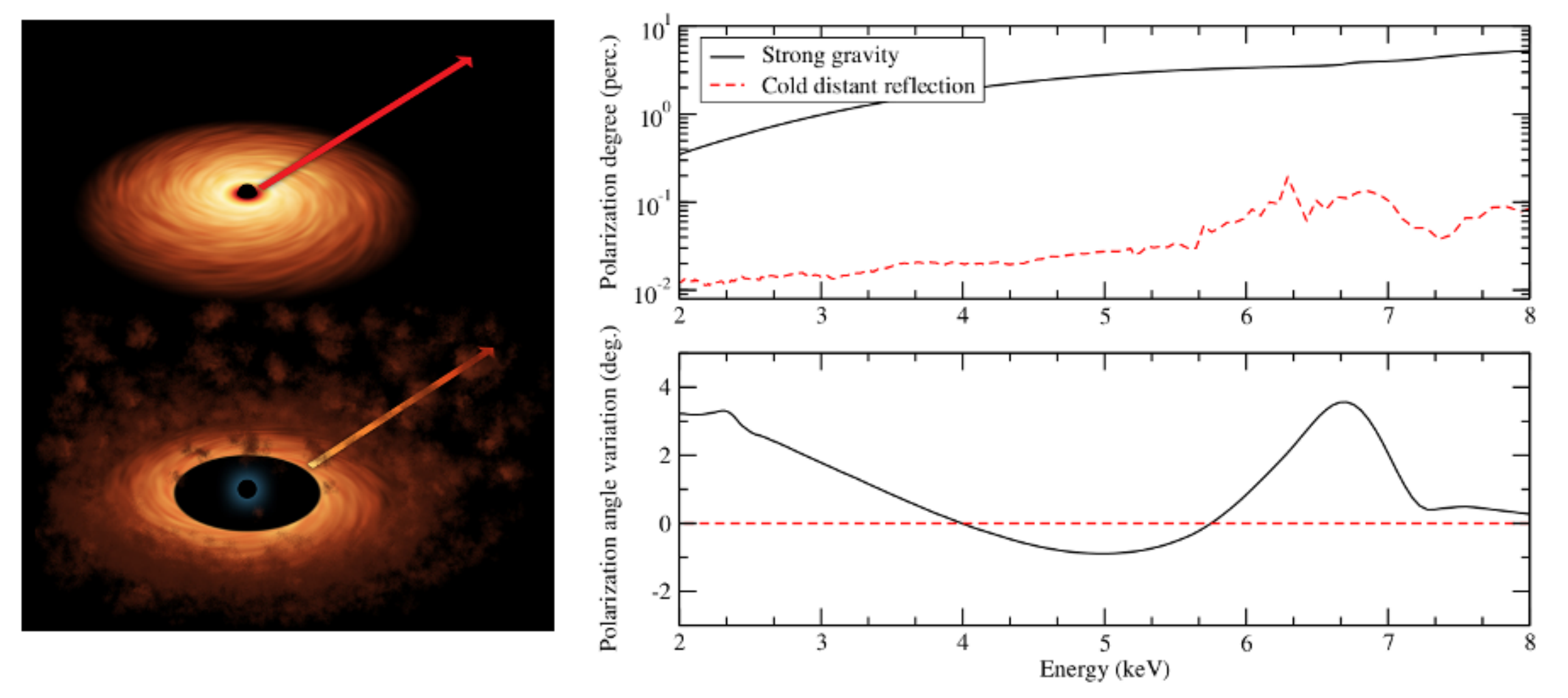}
  \caption{Two alternative models to explain the observed 
	   asymmetrical broadening of the iron K$\alpha$ line
	   in Seyfert-1 galaxies. The left-hand panel shows 
	   that the feature can be produced by special and 
	   general relativistic effects close to the potential
	   well (top) or by pure absorption and Compton scattering 
	   in a distant cloudy medium (bottom). Credits: 
	   NASA/JPL-Caltech. The right-hand panel is the 
	   polarization degree and polarization angle variation
	   associated with the two models in the IXPE energy 
	   band. Simulations from \citet{Marin2012}.}
  \label{marin:fig2}
\end{figure}

Another application of X-ray polarization measurements from AGN is the determination of the importance of Compton scattering 
in a distant cloudy medium onto the true shape of the relativistically-broadened iron K$\alpha$ line \citep{Marin2012}. 
If the broadening of the red wing of the emission line at 6.4~keV is not solely due to strong gravity effects near the black 
hole horizon, then the spin determined by the reflection method is probably over-estimated \citep{Miller2008}. Down-scattering 
of photons onto gaseous clumps along the observer's line-of-sight can significantly shift the line centroid, resulting in 
stronger asymmetries.

We show in Fig.~\ref{marin:fig2} that the two scenarios give very different polarization signatures in the IXPE band. In particular, 
the polarization degree of the gravity-dominated model is more than ten times stronger than the absorption scenario and, due to 
the energy-dependent albedo and scattering phase function of the disk material, the relativistic model shows a non-constant polarization 
position angle with energy. It is quite probable that both mechanisms are happening at the same time and X-ray polarization can 
definitively determine the dominant process.

\subsection{Magnetic fields strength in jets}

\begin{figure}[ht!]
 \centering
 \includegraphics[trim = 0mm 0mm 0mm 0mm, clip, width=9cm]{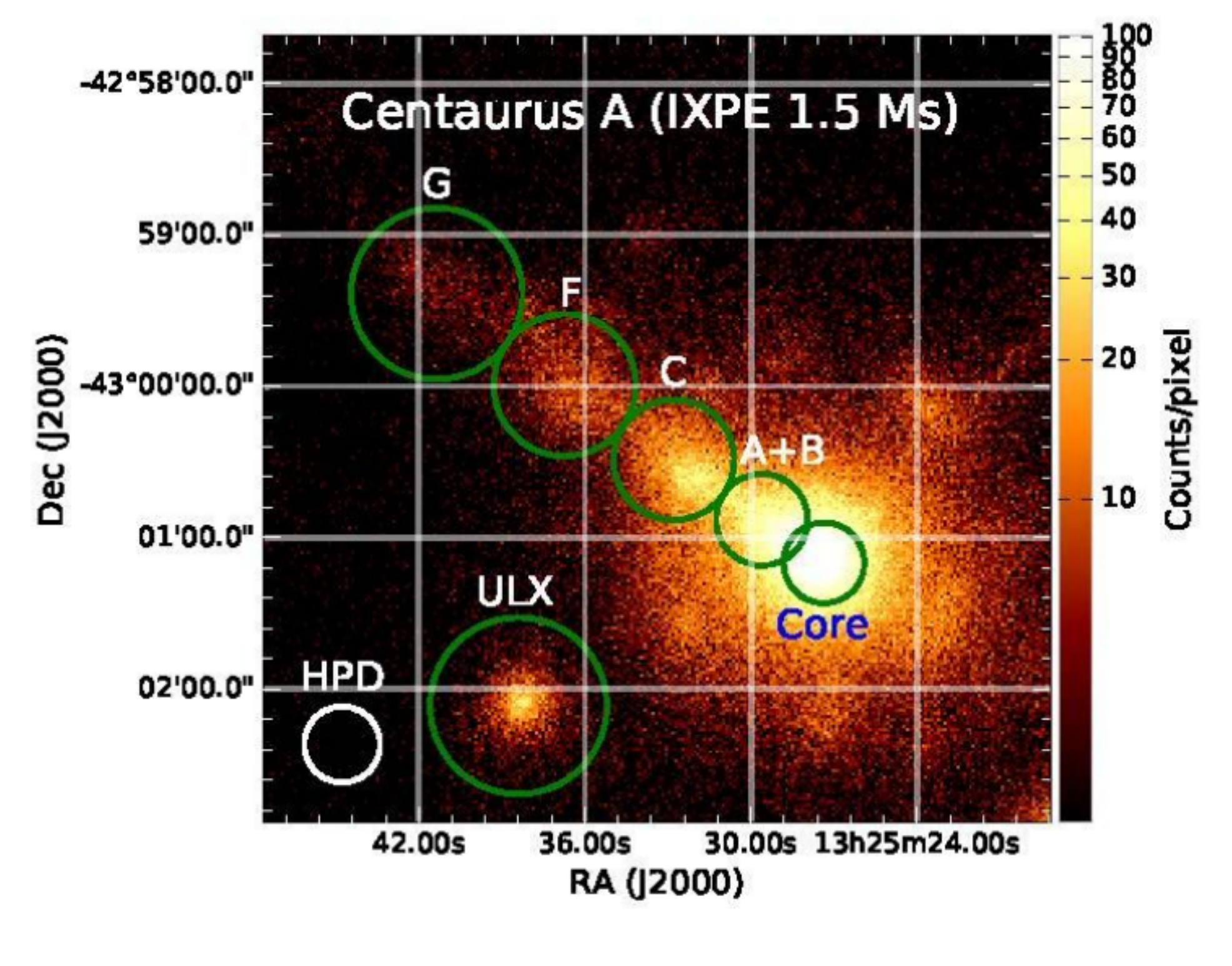}
  \caption{IXPE-convolved Chandra image of the X-ray brightest extragalactic 
	   source Centaurus~A (Cen~A). The white circle indicates the 
	   half-power diameter of IXPE and the green circles 
	   (angular resolution) demonstrate the possibility of 
	   the instrument to perform space-resolved X-ray polarization, 
	   testing the structure of the magnetic field along the jet. 
	   An ULX, in the field of view of Cen~A, can be 
	   simultaneously picked up. Credits: IXPE team.}
  \label{marin:fig3}
\end{figure}

Radio-loud AGN, whose spectral energy distribution is dominated by jet-induced synchrotron emission in the radio-band, are also 
excellent targets for IXPE. For a nearby radio galaxy such as Cen~A (4.6 Mpc and the X-ray brightest extragalactic source), the 
imaging capability of the satellite offers the possibility to perform space-resolved polarization studies, testing the structure 
of the magnetic field along the jet \citep{Weisskopf2016}. 

Fig.~\ref{marin:fig3} shows that is is possible with IXPE to map the magnetic field of resolved X-ray emitting jets close to 
the injection point of the electrons. Fig.~\ref{marin:fig3} is a convolved Chandra image of Cen~A with the IXPE response, together 
with a plausible model for the magnetic fields. The model consists of a transverse field in hot spots (shocks) along the jet,
a longitudinal field between hot spots and, in the core, a polarization of 30\% was assumed in order to estimate position-angle 
error \citep{McNamara2009}. Interestingly, the imaging capabilities of the instrument allow us to simultaneously pick up 
ultra-luminous X-ray sources in the field and one is show in Fig.~\ref{marin:fig3}.

\section{Conclusions}
IXPE will revolutionize our comprehension of the energetic Universe by opening a new observational window. The measurement of 
X-ray polarization and/or significantly small upper limits from a large variety of sources will enable us to test our numerical 
models by reducing their degrees of freedom. It is quite probable that many of our current theories will be revised in the light 
of IXPE observations. In this proceedings, we have shown three different results we expect from observations of AGN. Among others 
constraints, we expect to obtain precise measurements of the spin of black holes, together with a proper estimate of the impact 
of Compton down-scattering on the iron K$\alpha$ line profile. We will also target radio-loud, jet-dominated AGN in order to 
probe the structure and strength of their kilo-parsec scale magnetic fields.

\begin{acknowledgements}
F.M acknowledges funding through the CNES post-doctoral position grant 
``Probing the geometry and physics of active galactic nuclei with 
ultraviolet and X-ray polarized radiative transfer''. 
\end{acknowledgements}

\bibliographystyle{aa} 
\bibliography{marin} 

\end{document}